\documentclass[12pt]{article}
\usepackage{euscript}
\usepackage{amssymb}
\usepackage{amsmath}
\usepackage{graphicx}
\usepackage[small,sc,center]{titlesec}
\usepackage{indentfirst}

\newcommand{\nc}{\newcommand}
\nc{\ek}{E_\mathrm{K}}
\nc{\epsg}{\varepsilon_\mathrm{g,c}}
\nc{\epsn}{\varepsilon_\mathrm{n,c}}
\nc{\K}{\,\mathrm{K}}
\nc{\mzams}{M_\mathrm{ZAMS}}
\nc{\rph}{r_\mathrm{ph}}
\nc{\rphin}{r_{\mathrm{ph},0}}
\nc{\Teff}{T_\mathrm{eff}}
\nc{\tev}{t_\mathrm{ev}}

\begin{document}

\begin{center}
\textbf{EVOLUTION AND PULSATION PERIOD CHANGE IN THE LARGE MAGELLANIC CLOUD CEPHEIDS}

\textbf{Yu. A. Fadeyev\footnote{E--mail: fadeyev@inasan.ru}}

\textit{Institute of Astronomy, Russian Academy of Sciences, Pyatnitskaya ul. 48, Moscow, 109017 Russia}

Received June 3, 2013

\end{center}

\textbf{Abstract} ---
Theoretical estimates of the pulsation period change rates in LMC Cepheids
are obtained from consistent calculation of stellar evolution and nonlinear
stellar pulsation for stars with initial chemical composition $X=0.7$, $Z=0.008$,
initial masses $5M_\odot\le\mzams\le 9M_\odot$ and pulsation periods ranged
from 2.2 to 29 day.
The Cepheid hydrodynamical models correspond to the evolutionary stage
of thermonuclear core helium burning.
During evolution across the instability strip in the HR diagram
the pulsation period $\Pi$ of Cepheids is the quadratic function of the evolution
time for the both fundamental mode and first overtone.
Cepheids with initial masses $\mzams\ge 7M_\odot$ pulsate in the
fundamental mode and the period change rate $\dot\Pi$ varies
nearly by a factor of two for both crossings of the instability strip.
In the period -- period change rate diagram the values of $\Pi$ and $\dot\Pi$
concentrate within the strips, their slope and half--width depending
on both the direction of the movement in the HR--diagram and the pulsation mode.
For oscillations in the fundamental mode the half--widths of the strip are
$\delta\log\dot\Pi = 0.35$ and $\delta\log\dot\Pi = 0.2$
for the first and the secon crossings of the instability strip, respectively.
Results of computations are compared with observations of nearly
700 LMC Cepheids.
Within existing observational uncertainties of $\dot\Pi$
the theoretical dependences of the period change rate on the pulsation
period are in a good agreement with observations.


\section*{introduction}

Periods of light variations in many $\delta$~Cep pulsating type variables
(Cepheids) are known with eight significant digits (Samus et al. 2012).
So high accuracy of determination of the period $\Pi$ is due
excellent repetition of pulsation motions and also is owing to the fact that
photographic observations of Cepheids are carried out since the end of the
XIX century, so that photometric measurements of some stars of this type cover
as many as several thousands oscillation cycles.
In such a case the long--term observations allow us to significantly correct
the value of the period with the $O-C$ diagram.
At the same time as early as in the thirties of the XX century the $O-C$ diagrams
of some Cepheids were found to have the quadratic term indicating
the secular period change (Kukarkin and Florja 1932).
Interest in such a property grew after works by Hofmeister et al. (1964)
and Iben (1966) where the evolutionary state of Cepheids was determined
and long--term period changes were thought to be due to evolutionary changes
of the stellar structure during thermonuclear core helium burning
(Fernie 1979; Mahmoud and Szabados 1980; Szabados 1983; Deasy and Wayman 1985).

In recent years a great deal of observational data on long--term pulsation period
changes in the Large Magellanic Cloud (LMC) Cepheids was obtained in
ASAS, MACHO and OGLE projects.
Pietrukowicz (2001) considered data on 378 LMC Cepheids and concluded that all
studied variables with periods longer 8 days show period changes.
Later Poleski (2008) carried out an analysis of 655 LMC Cepheids and found that
18\% of fundamental mode and 41\% of first overtone pulsators have
evolutionary period changes.

The estimation of the period change rate $\dot\Pi$ from observations
is of great interest since it provides with the direct test of the stellar
evolution theory.
Unfortunately, theoretical studies of pulsation period changes in Cepheids
based on consistent solution of the equations of stellar evolution and
stellar pulsation have not been done yet.
Pietrukowicz (2001) compared his observational data with evolutionary and
pulsation models studied by Alibert et al. (1999) and Bono et al. (2000).
However the pulsation period change rates $\dot\Pi$ were not evaluated in these
theoretical works, so that Pietrukowicz (2001) used rough estimates from
presented tabular data.
Moreover, Alibert et al. (1999) in their linear analysis of pulsational
instability did not take into account effects of convection.
Such a simplification might be responsible for a large disagreement between
theoretical models and observational estimates of $\dot\Pi$ (Pietrukowicz 2001).
Poleski (2008) compared his observational data with stellar evolution
theory using the approach by Turner et al. (2006) which is also based on
strong simplifications.

The goal of the present work is to obtain theoretical estimates of the
pulsation period change rate $\dot\Pi$ as a function of the age of the Cepheid
using the consistent calculations of stellar evolution and nonlinear stellar
pulsation.
Initial relative mass abundances of hydrogen and elements heavier than helium
correspond to the LMC chemical composition: $X=0.7$, $Z=0.008$.
In hydrodynamical calculations of nonlinear stellar pulsations we take into
account effects of turbulent convection, so that the hydrodynamical models
occupy the whole interval of effective temperatures bounded in the
Hertzsprung--Russel (HR) diagram  by the blue and red edges of
the instability strip.
Methods of stellar evolution calculation and basic equations of radiation
hydrodynamics and turbulent convection used for calculation of
nonlinear stellar pulsation are given in our previous paper (Fadeyev 2013).

\section*{results of computations}

Solution of the equations of hydrodynamics for nonlinear stellar oscillations
as a function of time $t$ was done with initial conditions taken in the form of
stellar models of evolutionary sequences of stars with initial masses
$5M_\odot\le\mzams\le 9M_\odot$.
Evolutionary tracks in the HR diagram of stars under consideration
are shown in Fig.~1 where in dotted lines are shown parts of the track
corresponding to the instability against radial oscillations.
In the starting and in the ending track points the rate
of the thermonuclear energy generation rate $\epsn$ and the rate of the gravitational
energy production $\epsg$ in the stellar center are nearly the same:
$\epsn\approx\epsg$.
Therefore the tracks displayed in Fig.~1 represent the evolutionary stage
when the only source of energy generation in the stellar center
is thermonuclear helium burning.

For each evolutionary track the bounds of pulsational instability in the HR diagram
were determined from hydrodynamical computations where as in our previous work
(Fadeyev 2013) the kinetic energy of pulsation motions $\ek$ was calculated.
The part of the evolutionary track with pulsational instability was
determined from condition $\eta > 0$, where
$\eta = \Pi^{-1}d\ln{\ek}_{\max}/dt$ is the growth rate of the kinetic energy,
${\ek}_{\max}$ is the maximum value of the kinetic energy reached during
one pulsational cycle.
The pulsation period $\Pi$ was evaluated from the discrete Fourier transform
of the kinetic energy $\ek$.
It should be noted that the interval of time $t$
within of which we integrate equations of hydrodynamics is
comparable with the thermal scale of outer layers of the Cepheid
and is much shorter in comparison with the nuclear evolution time scale.
Fot example, in the Cepheid with initial mass $\mzams=7M_\odot$
the evolution time between the red and blue edges of the instability strip
is $\sim 10^5$~years, whereas hydrodynamic computations of the instability growth
with subsequent limit cycle attainment are done on the time interval of $\sim 10$~years.

In Fig.~2 we give the plots of the instability growth rate $\eta$
versus effective temperature averaged over the pulsational cycle
$\langle\Teff\rangle$ for four Cepheid evolutionary sequences with
initial masses form $5M_\odot$ to $8M_\odot$.
The evolutionary track crosses the instability strip twice in the HR diagram
and therefore each evolutionary sequence is represented by two plots
where the first one correponds to the movement across the HR diagram with
increasing effective temperature (dotted lines) and the second plot
corresponds to the movement in the opposite direction (dash--dotted lines).
In Fig.~2 we use the averaged over the cycle effective temperature $\langle\Teff\rangle$
as independent variable because in the hydrodynamical model
the average radius of the photosphere $\langle\rph\rangle$ is
smaller than the radius of the photosphere of the hydrostatically equilibrium model
$\rphin$.
For hydrodynamical models of Cepheids calculated in the present study
the ratio of the photosphere radii ranges within
$0.975\le\langle\rph\rangle/\rphin < 1$
and edges of the instability strip shift to the blue in the HR diagram by
$30\K < \Delta\Teff < 70\K$.

The presence of two maxima in plots of $\eta$ for $\mzams\le 6M_\odot$
is due to the fact that near the red edge of the instability strip
radial pulsations are excited in the fundamental mode, whereas at higher
effective temperatures pulsations are excited in the first overtone.
Transition between oscillation modes takes place within the effective temperature
range $6000\K <\langle\Teff\rangle < 6100\K$.

In Cepheids with initial mass $\mzams = 7M_\odot$ evolving blueward
across the instability strip radial oscillations are due to
instability of the fundamental mode and transition to the first overtone
takes place just near the blue edge at $\langle\Teff\rangle\approx 6200\K$.
During the second crossing of the instability strip radial oscillations exist
in the form of the fundamental mode.

Pulsations of Cepheids with initial mass $\mzams\ge 8M_\odot$ are always due to
instability of the fundamental mode and the pulsation period $\Pi$ gradually
changes while the star moves in the HR diagram from one edge of the instability
strip to another.
The change of the pulsation period of the Cepheid with initial mass $\mzams = 8M_\odot$
is illustrated in Fig.~3 where for the sake of convenience we set the evolution time
$\tev$ to zero when the star crosses the edge of the instability strip and
begins to oscillate.
The plot with gradual decrease of the pulsation period corresponds to the first
crossing of the instability strip and the plot with gradually increasing period
corresponds to the second crossing.
Hydrodynamical models with positive and negative growth rates $\eta$ are shown
in filled circles and opened circles, respectively.
As is seen from shown plots the pulsation period $\Pi$ is fitted
by an algebraic polynomial $\Pi(\tev) = a_0 + a_1\tev + a_2 \tev^2$
for both evolutionary sequences
with a good accuracy (i.e. with relative r.m.s. error less than one per cent).
Polynomial approximation is shown in Fig.~3 by dotted and dash--dotted lines
for the first crossing and the second crossing of the instability strip, respectively.

Expression of the pulsation period $\Pi$ as a quadratic polynomial of $\tev$
was found to be a good approximation for all Cepheid models considered
in the present study.
The only exception is a discontinuity of the period due to transition
from one pulsation mode to other.
This is illustrated in Fig.~4 by the plots of the pulsation period for the
Cepheid with initial mass $\mzams = 6M_\odot$.
However within the interval of the continuous change of $\Pi$
the quadratic polynomial remains a quite good approximation.

The quadratic dependence of the pulsation period $\Pi$ on the evolutionary time
$\tev$ implies the linear change of $\dot\Pi$ which decreases during the first
crossing of the instability strip and increases during the next crossing.
In Cepheids pulsating in the fundamental mode within the whole instability strip
the period change rate $\dot\Pi$ varies roughly by a factor of two.
Typical values of $\dot\Pi$ can be found in the table where for the evolutionary sequences
with initial masses $5M_\odot\le\mzams\le 9M_\odot$
we give the main properties of Cepheids at the points where the evolutionary track
crosses the edges of the instability strip.
Each evolutionary sequence is represented by four lines where the first
pair of lines corresponds to the first crossing of the instability strip
and the second pair of lines corresponds to the second crossing.
In the second column of the table we give the evolution time $\Delta\tev$
spent by the Cepheid within instability strip.
In following columns we give main parameters of the Cepheid at the edge
of the instability strip (i.e. for $\eta = 0$)
which were obtained by linear interpolation of model parameters
of adjacent hydrodynamical models with opposite signs of the growth rate
of kinetic energy.
The model parameters are as follows:
the stellar mass $M$ which is less than the initial mass $\mzams$ due to effects
of the stellar wind during the preceding evolution;
the averaged over the cycle absolute bolometric luminosity $L$
and effective temperature $\langle\Teff\rangle$;
the pulsation period $\Pi$,
the dimensionless pulsation period change rate $\dot\Pi$;
the order of the pulsation mode $k$ ($k=0$ for the fundamental mode and
$k=1$ for the first overtone).

\section*{comparison with observations}

The period of light variations is the only quantity which can be determined
from observations of the pulsating variable star with sufficiently high precision.
Therefore for comparison of the results of theoretical computations with
observational data we will consider the period change rate $\dot\Pi$
as a function of the pulsation period $\Pi$.

In Fig.~5 the plots of the dimensionless period change rate $\dot\Pi$ are shown
as a function of the pulsation period $\Pi$ for the Cepheid models crossing
blueward the instability strip.
The plots are separated into two groups with Cepheids pulsating in the
fundamental mode ($6.9~\mathrm{day}\le\Pi\le 28~\mathrm{day}$) and those
pulsating in the first overtone ($2.2~\mathrm{day}\le\Pi\le 4.5~\mathrm{day}$).
The plots locate along the dashed lines which are approximately given by
following relations
\begin{equation}
\label{p-pdot1}
\log(-\dot\Pi) = \left\{
\begin{array}{ll}
-11.71 + 4.836 \log\Pi , \qquad & k=0 , \\
-9.676 + 2.562 \log\Pi ,        & k=1 ,
\end{array}
\right.
\end{equation}
where the period $\Pi$ is expressed in days, whereas $k=0$ and $k=1$ correspond to the
fundamental mode and to the first overtone, respectively.
Due to the finite width of the instability strip the plots of evolutionary sequences
shown in Fig.~5 by solid lines are confined within the bands with half--width
$\delta\log\dot\Pi = 0.035$ for $k=0$
and $\delta\log\dot\Pi\approx 0.1$ for $k=1$.

The diagram period -- period change rate for Cepheids of the second crossing
of the instability strip is shown in Fig.~6.
Unfortunately, reliable estimates of the period change rate for Cepheid models
with initial mass $\mzams = 5M_\odot$ evolving redward in the HR diagram
were not obtained, so that in Fig.~6 the mean dependence of the period change rate
is shown only for the fundamental mode:
\begin{equation}
\label{p-pdot2}
\log\dot\Pi = -10.33 + 3.386 \log\Pi .
\end{equation}
The half--width of the band in the period -- period change rate diagram is
$\delta\log\dot\Pi = 0.2$.

To compare results of our theoretical computations with observations we used
observational estimates of the period $\Pi$ and the period change rate $\dot\Pi$
from works by Pietrukowicz (2001) and Poleski (2008).
Electronic tables 1 -- 3 supplementing the paper by Pietrukowicz (2001)
give data on 369 LMC Cepheids, whereas the period change rates were evaluated
using the Harvard photographic observations obtained in time interval from 1910 to 1950.
Observational data on LMC Cepheids obtained from the OGLE survey were
received from the author (Poleski 2008).

The diagram period -- period change rate for Cepheids crossing the instability strip
blueward with negative $\dot\Pi$ is presented in Fig.~7a
and the same diagram for Cepheids evolving redward with positive $\dot\Pi$
is shown in Fig.~7b.
Results of observations obtained by Pietrukowicz (2001) and Poleski (2008)
are shown by filled circles and open circles, respectively.
Theoretical dependences obtained in the present study for Cepheid evolutionary
sequences with initial masses from 5 to $9M_\odot$ are shown in solid lines.
In general one can conclude that the theory of stellar evolution 
agrees with observations of Cepheids.

At the same time one should note a disagreement between observational
results by Pietrukowicz (2001) and those by Poleski (2008).
A possible cause of such a difference seems to be a shorter time interval
used by Poleski (2008) for evaluation of the Cepheid period changes.

\section*{conclusion}

Results of our calculations allow us to conclude that the oscillation period $\Pi$
of the Cepheid is the quadratic function of the evolution time $\tev$
and when the star crosses the instability strip the quantity $\dot\Pi$
changes by a factor of two.
Earlier Deasy and Wayman (1985) noted the occurence of non--constant period change
in LMC Cepheids.

An interesting result of our calculations is that the dependence of the
period change rate $\dot\Pi$ on the pulsation period $\Pi$ for the
fundamental mode differs from that for the first overtone.
Unfortunately, at present observational confirmation of this feature seems to be
impossible because of insufficiently high accuracy of observational evaluation
of $\dot\Pi$ for the first overtone Cepheids.
Significant scatter of points in Fig.~7 at short periods ($\Pi < 7$~day)
is mainly due to the power law decrease of $\dot\Pi$ with decreasing pulsation period.
Indeed, as is seen in Fig.~5 the period change rates in first overtone Cepheids
are three orders of magnitude smaller in comparison with those in long period
($\Pi\approx 30$~day) Cepheids.
Therefore, to reduce the error of the observational estimate of $\dot\Pi$
in first overtone pulsators
to the value comparable with that in long period Cepheids
the time interval of the $O-C$ diagram should be expanded by two orders
of magnitude.

Thus, for comparison of the stellar evolution calculations
with observations of most interest are fundamental mode Cepheids.
Here among the important questions we should emphasize
the role of chemical composition and convective overshooting in the
period -- period change rate diagram.
It should also be noted that the crossing time of the Cepheid instability strip 
by core helium burning stars with initial masses $\mzams > 9M_\odot$ becomes
comparable with that
during the gravitational contraction of the helium core before
the stage of the red supergiant.
Thefore the most massive long period Cepheids with gravitationally contracting core
may play perceptible role in the period -- period change rate diagram.

The author thanks Radek Poleski who kindly placed the Cepheid OGLE data
at his disposal.
The study was supported by the Basic Research Program of the Russian Academy of Sciences
``Nonstationary phenomena in the Universe''.

\subsection*{REFERENCES}

\begin{enumerate}
\item Y. Alibert, I. Baraffe, P. Hauschildt, et al., Astron.Astrophys. \textbf{344}, 551 (1999).

\item G. Bono, F. Caputo, S. Cassisi, et al., Astrophys.J. \textbf{543}, 955 (2000).

\item H.P. Deasy and P.A. Wayman, MNRAS \textbf{212}, 395 (1985).

\item Yu.A. Fadeyev, Pis'ma Astron. Zh. \textbf{39}, 342 (2013)
      [Astron.Lett. \textbf{39}, 306 (2013)].

\item J.D. Fernie, Astrophys. J. \textbf{231}, 841 (1979).

\item E. Hofmeister, R. Kippenhahn and A. Weigert, Zeitschrift f\"ur Astrophys. \textbf{60}, 57 (1964).

\item I. Iben, Astrophys.J. \textbf{143}, 483 (1966).

\item B.W. Kukarkin and N. Florja, Zeitschrift f\"ur Astrophys. \textbf{4}, 247 (1932).

\item F. Mahmoud and L. Szabados, IBVS, N 1895, 1 (1980).

\item P. Pietrukowicz, Acta Astron. \textbf{51}, 247 (2001).

\item R. Poleski, Acta Astron. \textbf{58}, 313 (2008).

\item N.N. Samus, O.V. Durlevich, E.V. Kazarovets, et al.,
      \textit{General Catalogue of Variable Stars} (GCVS database, version April 2012),
      CDS B/gcvs (2012).

\item L. Szabados, Astrophys. Space Sci \textbf{96}, 185 (1983).

\item D. Turner, M. Abdel--Sabour Abdel--Latif, and L.N. Berdnikov, PASP \textbf{118}, 410 (2006).

\end{enumerate}

\newpage
\centerline{Cepheid models at the edges of the instability strip}

\vskip 15pt
\begin{tabular}{l|l|r|r|r|r|r|r}
\hline
 $\mzams/M_\odot$ & $\log\Delta\tev$, year & $M/M_\odot$ & $L/L_\odot$, $10^4$ & $\langle\Teff\rangle$, K & $\Pi$, day & $\dot\Pi$, $10^{-7}$ & $k$\\
\hline
   9    & 3.57 &   8.715  &   1.4081  &   5467  &   25.62  &  -70.3   &  0  \\
        &      &   8.715  &   1.4259  &   6012  &   18.17  &  -39.1   &  0  \\
        & 3.89 &   8.664  &   1.5440  &   5905  &   20.98  &   16.0   &  0  \\
        &      &   8.664  &   1.5106  &   5458  &   27.59  &   30.5   &  0  \\[4pt]
   8.5  & 3.77 &   8.262  &   1.1553  &   5475  &   22.13  &  -42.9   &  0  \\
        &      &   8.262  &   1.1718  &   6079  &   15.36  &  -19.3   &  0  \\
        & 4.12 &   8.216  &   1.3126  &   6025  &   17.45  &    9.74  &  0  \\
        &      &   8.215  &   1.2784  &   5464  &   24.58  &   19.7   &  0  \\[4pt]
   8    & 3.98 &   7.806  &   0.9313  &   5510  &   18.58  &  -23.0   &  0  \\
        &      &   7.806  &   0.9454  &   6134  &   12.78  &  -10.5   &  0  \\
        & 4.31 &   7.764  &   1.0937  &   6092  &   14.88  &    5.76  &  0  \\
        &      &   7.763  &   1.0615  &   5459  &   21.68  &   12.6   &  0  \\[4pt]
   7    & 4.88 &   6.871  &   0.5775  &   5548  &   12.95  &  -3.84   &  0  \\
        &      &   6.869  &   0.5937  &   6303  &    5.70  &  -0.347  &  1  \\
        & 4.73 &   6.831  &   0.7204  &   6255  &   10.31  &   1.72   &  0  \\
        &      &   6.829  &   0.6963  &   5512  &   15.75  &   3.99   &  0  \\[4pt]
   6    & 5.71 &   5.917  &   0.3436  &   5635  &    8.70  &  -0.295  &  0  \\
        &      &   5.913  &   0.3716  &   6548  &    3.76  &  -0.0699 &  1  \\
        & 5.31 &   5.889  &   0.4285  &   6444  &    4.46  &   0.219  &  1  \\
        &      &   5.886  &   0.4129  &   5595  &   10.54  &   1.15   &  0  \\[4pt]
   5    & 6.09 &   4.955  &   0.1874  &   5796  &    5.26  &  -0.0604 &  0  \\
        &      &   4.952  &   0.2096  &   6808  &    2.31  &  -0.0193 &  1  \\
        & 6.20 &   4.947  &   0.2208  &   6771  &    2.46  &   0.0174 &  1  \\
        &      &   4.942  &   0.2178  &   5751  &    6.22  &   0.154  &  0  \\
\hline
\end{tabular}
\clearpage

\newpage
\section*{FIGURE CAPTIONS}

\begin{itemize}
\item[Fig. 1.] Evolutionary tracks of the core helium burning stars in the HR diagram.
               The initial stellar mass $\mzams$ is indicated near each track.
               Parts of tracks corresponding to the instability of the star against
               radial oscillations are shown by dotted lines.

\item[Fig. 2.] The kinetic energy growth rate $\eta$ versus the mean effective temperature
               $\langle\Teff\rangle$ of the star evolving across the Cepheid instability strip.
               The dotted and dash--dotted lines correspond to the blueward and redward
               evolution, respectively.
               Each pair of plots is arbitrarily shifted along the vertical axis
               and the horizontal dashed line indicates $\eta = 0$.
               Initial stellar masses $\mzams$ are indicated near the plots.

\item[Fig. 3.] The period of radial oscillations $\Pi$ of the Cepheid with initial mass
               $\mzams = 8M_\odot$ as a function of the evolution time $\tev$
               counted from the moment when the star enters the instability strip.
               Hydrodynamical models are shown by filled circles ($\eta > 0$) and
               open circles ($\eta < 0$).
               The second--order algebraic polynomial approximation is shown in
               dotted (blueward evolution in the HR diagram) and
               dash--dotted (redward evolution) lines.

\item[Fig. 4.] Same as Fig.~3 but for $\mzams = 6M_\odot$.
               Plots with periods $\Pi > 6$~day and $\Pi < 6$~сут
               correspond to radial pulsations in the fundamental mode
               and the first overtone, respectively.

\item[Fig. 5.] The dimensionless period change rate $\dot\Pi$ as a function of the
               pulsation period $\Pi$ for Cepheids during the first crossing of the
               instability strip.
               Relations (\ref{p-pdot1}) are shown in dashed lines for oscillations
               in the fundamental mode ($\Pi > 6.9$~day) and the first overtone
               ($\Pi < 4.5$~day).

\item[Fig. 6.] Same as Fig.~5 but for Cepheids during the second crossing
               of the instability strip.

\item[Fig. 7.] The dimensionless period change rate $\dot\Pi$ versus the pulsation period $\Pi$
               (in days) for Cepheids during the first (a) and the second (b) crossings of
               the instability strip.
               Observational data by Pietrukowicz (2001) and Poleski (2008) are shown
               in filled circles and open circles, respectively.
               Results of theoretical computations are shown in solid lines.
               Initial stellar masses are indicated at the curves.
               The unlabelled curve corresponds to the first overtone Cepheids with
               $\mzams = 6M_\odot$.

\end{itemize}

\newpage
\begin{figure}
\centerline{\includegraphics[width=15cm]{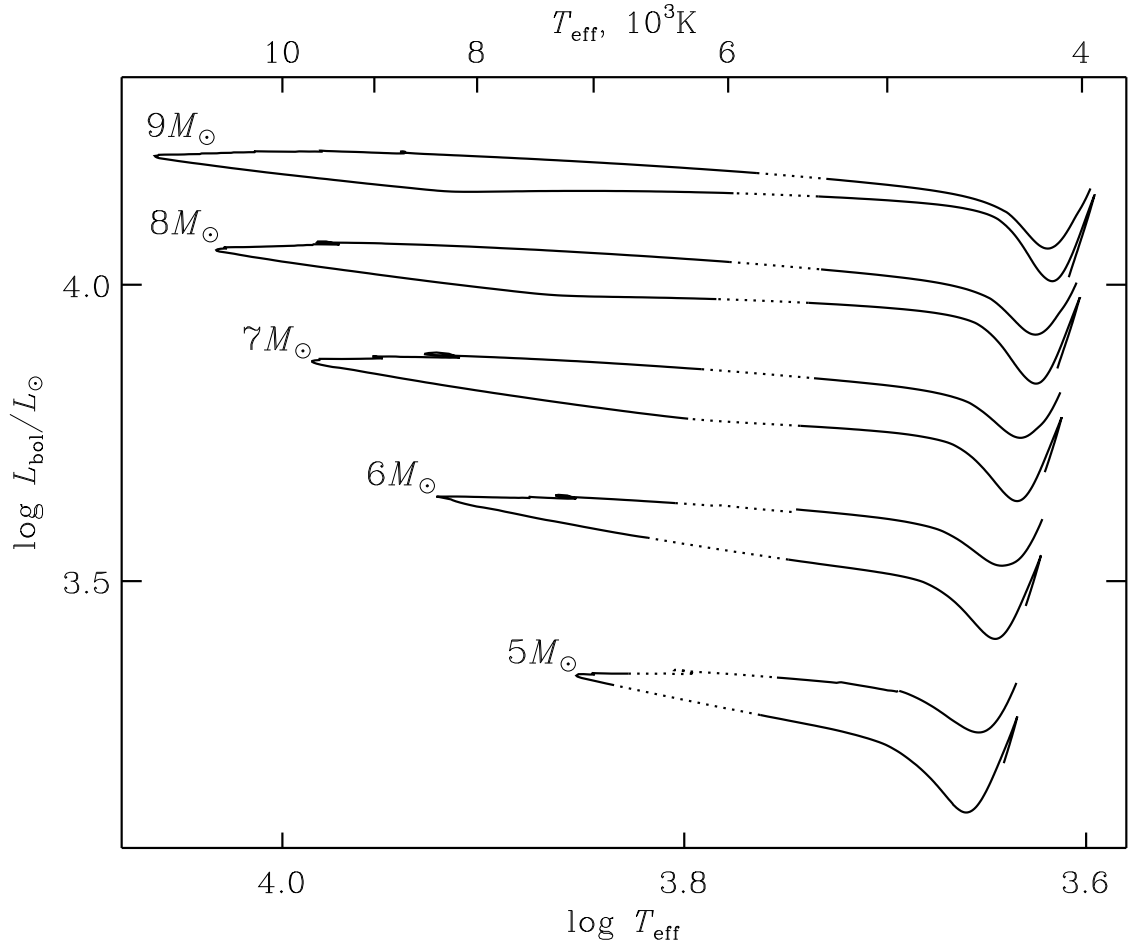}}
\caption{Evolutionary tracks of the core helium burning stars in the HR diagram.
         The initial stellar mass $\mzams$ is indicated near each track.
         Parts of tracks corresponding to the instability of the star against
         radial oscillations are shown by dotted lines.}
\label{fig1}
\end{figure}

\newpage
\begin{figure}
\centerline{\includegraphics[width=15cm]{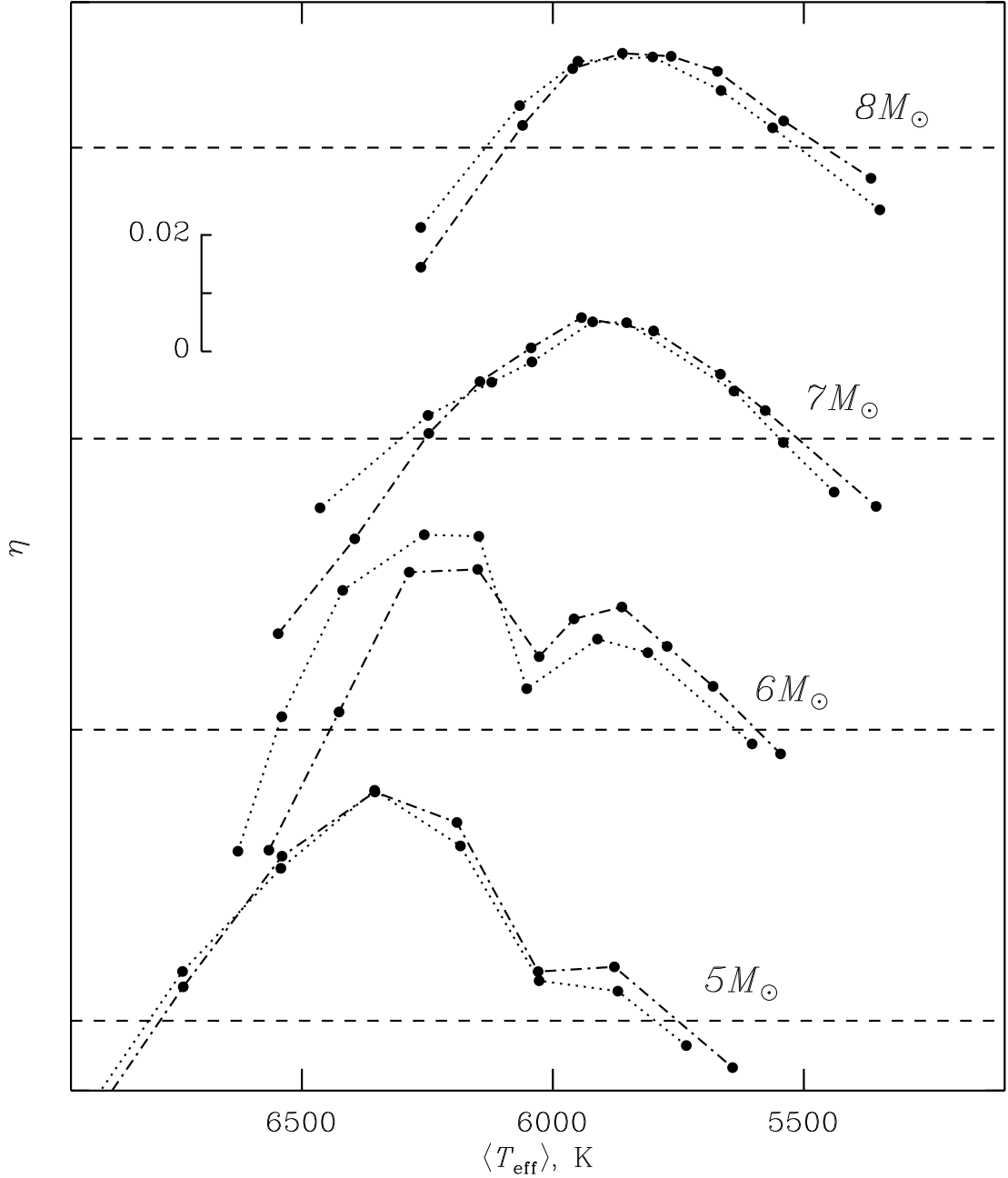}}
\caption{The kinetic energy growth rate $\eta$ versus the mean effective temperature
         $\langle\Teff\rangle$ of the star evolving across the Cepheid instability strip.
         The dotted and dash--dotted lines correspond to the blueward and redward
         evolution, respectively.
         Each pair of plots is arbitrarily shifted along the vertical axis
         and the horizontal dashed line indicates $\eta = 0$.
         Initial stellar masses $\mzams$ are indicated near the plots.}
\label{fig2}
\end{figure}

\newpage
\begin{figure}
\centerline{\includegraphics[width=15cm]{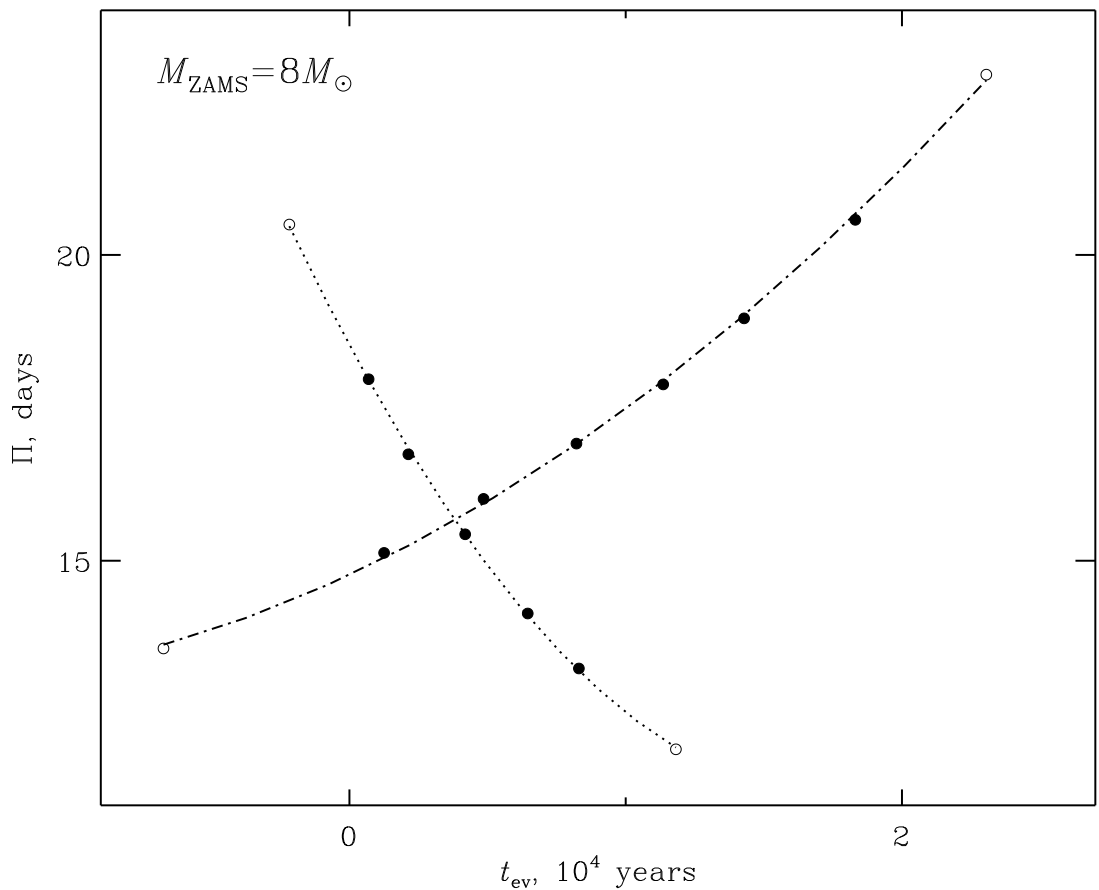}}
\caption{The period of radial oscillations $\Pi$ of the Cepheid with initial mass
         $\mzams = 8M_\odot$ as a function of the evolution time $\tev$
         counted from the moment when the star enters the instability strip.
         Hydrodynamical models are shown by filled circles ($\eta > 0$) and
         open circles ($\eta < 0$).
         The second--order algebraic polynomial approximation is shown in
         dotted (blueward evolution in the HR diagram) and
         dash--dotted (redward evolution) lines.}
\label{fig3}
\end{figure}

\newpage
\begin{figure}
\centerline{\includegraphics[width=15cm]{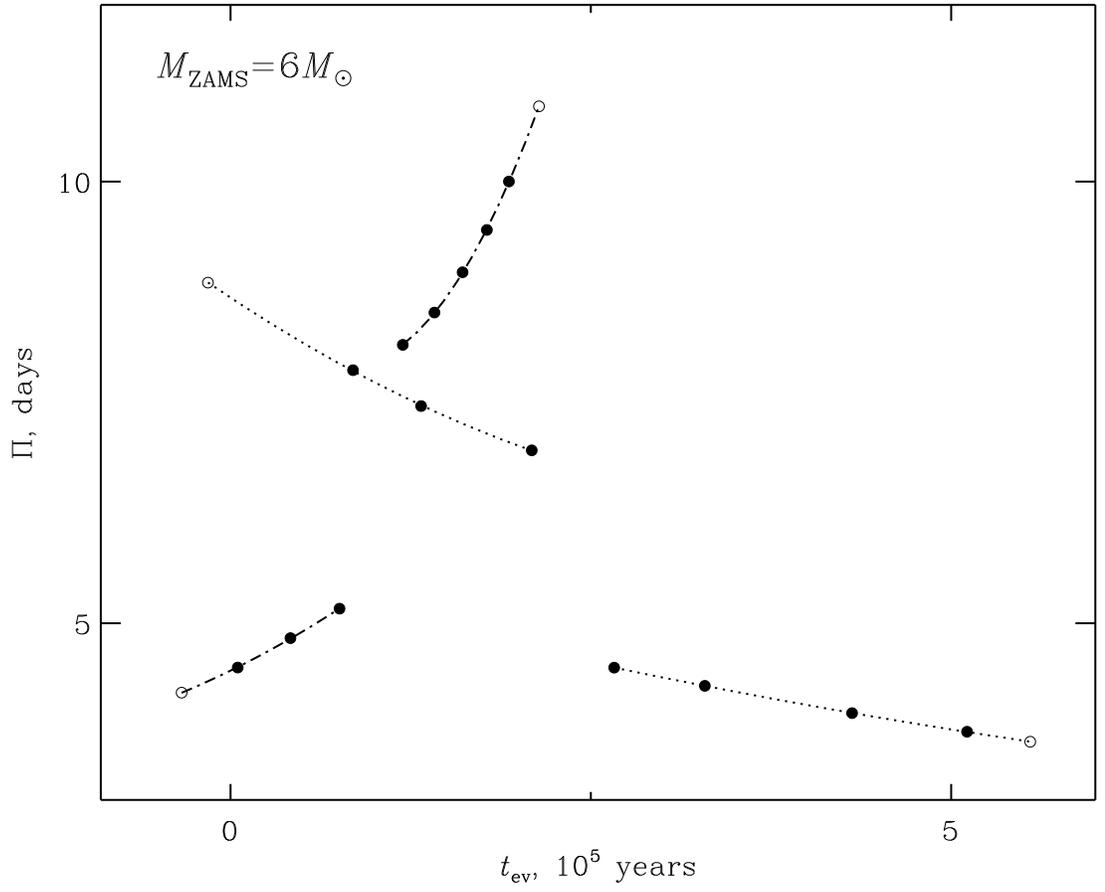}}
\caption{Same as Fig.~3 but for $\mzams = 6M_\odot$.
         Plots with periods $\Pi > 6$~day and $\Pi < 6$~сут
         correspond to radial pulsations in the fundamental mode
         and the first overtone, respectively.}
\label{fig4}
\end{figure}

\newpage
\begin{figure}
\centerline{\includegraphics[width=15cm]{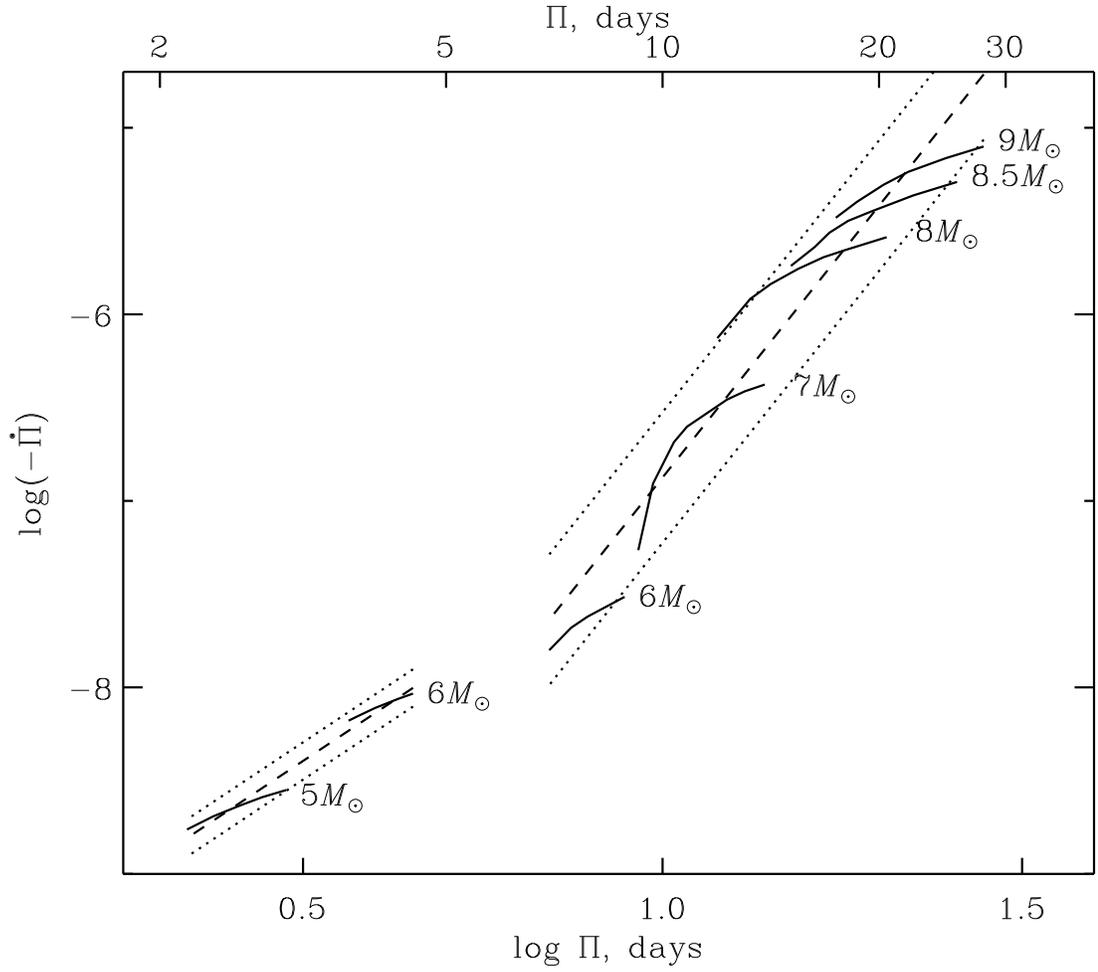}}
\caption{The dimensionless period change rate $\dot\Pi$ as a function of the
         pulsation period $\Pi$ for Cepheids during the first crossing of the
         instability strip.
         Relations (\ref{p-pdot1}) are shown in dashed lines for oscillations
         in the fundamental mode ($\Pi > 6.9$~day) and the first overtone
         ($\Pi < 4.5$~day).}
\label{fig5}
\end{figure}

\newpage
\begin{figure}
\centerline{\includegraphics[width=15cm]{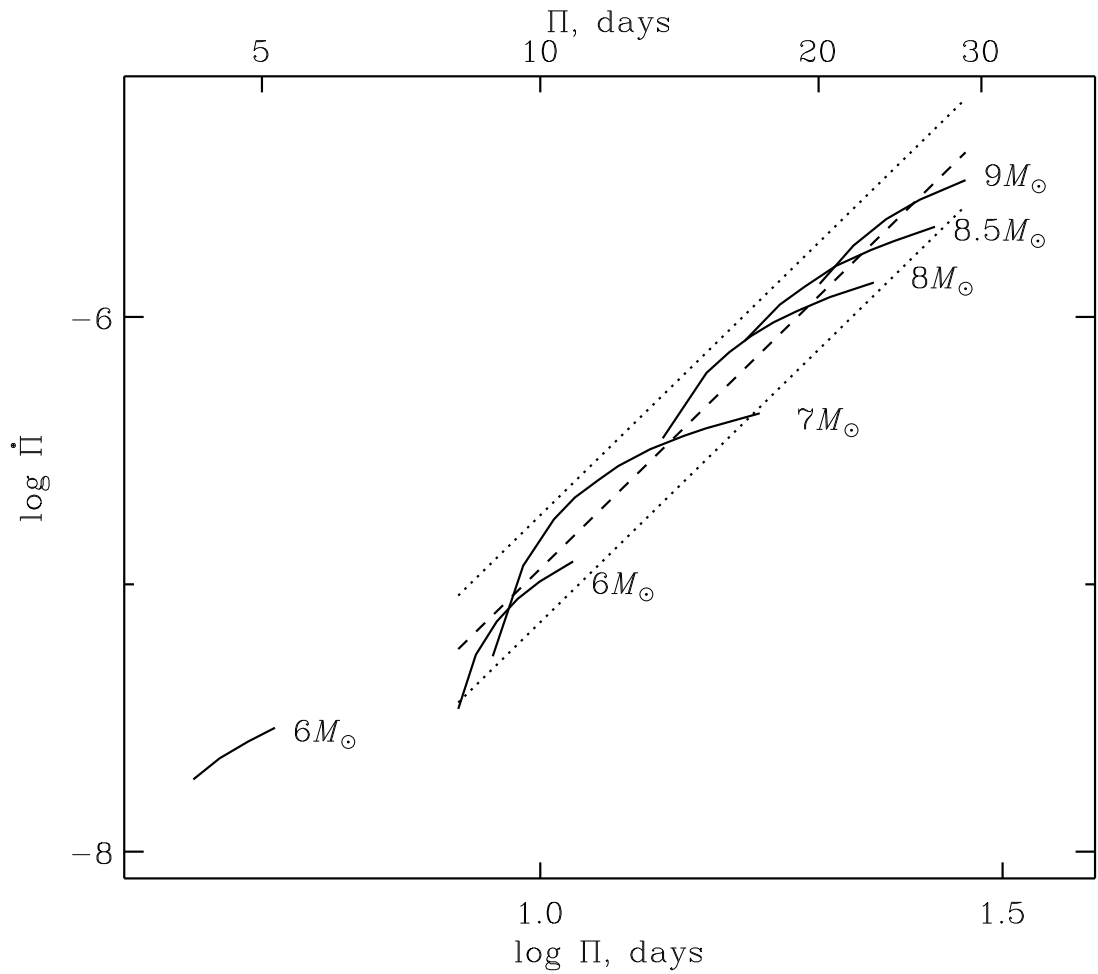}}
\caption{Same as Fig.~5 but for Cepheids at the second crossing of the instability strip.}
\label{fig6}
\end{figure}

\newpage
\begin{figure}
\centerline{\includegraphics[width=15cm]{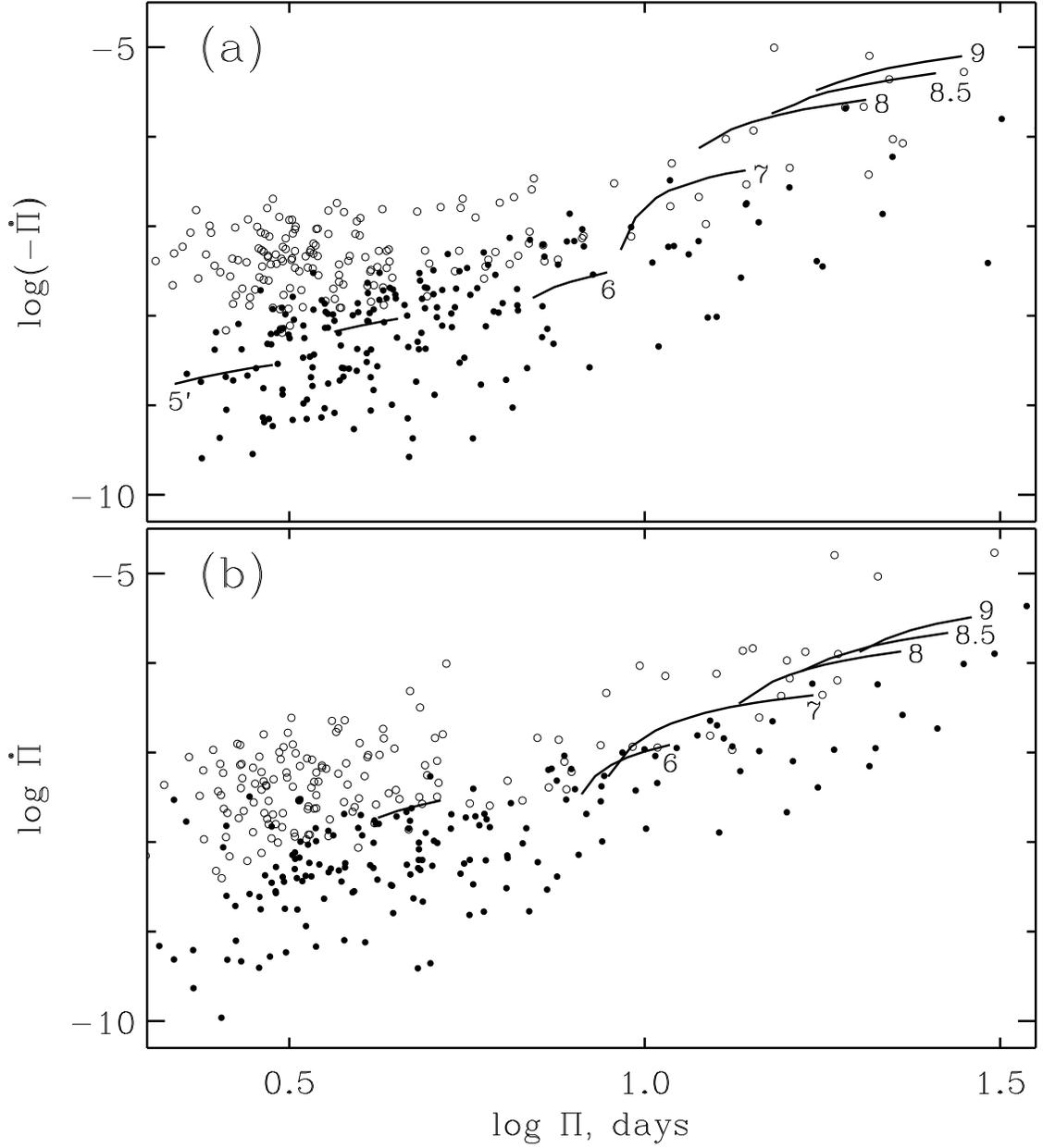}}
\caption{The dimensionless period change rate $\dot\Pi$ versus the pulsation period $\Pi$
         (in days) for Cepheids during the first (a) and the second (b) crossings of
         the instability strip.
         Observational data by Pietrukowicz (2001) and Poleski (2008) are shown
         in filled circles and open circles, respectively.
         Results of theoretical computations are shown in solid lines.
         Initial stellar masses are indicated at the curves.
         The unlabelled curve corresponds to the first overtone Cepheids with
         $\mzams = 6M_\odot$.}
\label{fig7}
\end{figure}

\end{document}